\documentclass{ws-ijgmmp}
\numberwithin{equation}{section}
\begin{document}
\title{SNUB 24-CELL DERIVED FROM THE COXETER--WEYL GROUP ${\bf W(D_4)}$ }
 
\author{Mehmet Koca\footnote{electronic-mail:kocam@squ.edu.om}, 
Nazife Ozdes Koca\footnote{electronic-mail:nazife@squ.edu.om} and 
Muataz Al-Barwani\footnote{electronic-mail:muataz@squ.edu.om}}

\address{Department of Physics, College of Science,
Sultan Qaboos University \\
P. O. Box 36, Al-Khoud, 123 Muscat,
Sultanate of Oman}

\maketitle

\begin{abstract}
Snub 24-cell is the unique uniform chiral polytope in four dimensions
consisting of 24 icosahedral and 120 tetrahedral cells. The vertices of
the 4-dimensional semi-regular polytope snub 24-cell and its symmetry
group $(W(D_4)/C_2):S_3$ of order 576 are obtained from the quaternionic representation of the
Coxeter-Weyl group  $W(D_4)$.The symmetry group is an extension of the proper subgroup of
the Coxeter-Weyl group $W(D_4)$ by the permutation symmetry of the Coxeter-Dynkin diagram $D_4$. 
The 96 vertices of the snub 24-cell are obtained as the orbit of the
group when it acts on the vector  $\Lambda = (\tau,1,\tau,\tau)$ or on the vector  $\Lambda = (\sigma,1,\sigma,\sigma)$
in the Dynkin basis with  $\tau=\frac{1+\sqrt{5}}{2} $ and $\sigma=\frac{1-\sqrt{5}}{2}  $.
  The two different sets represent the mirror images of the snub
24-cell. When two mirror images are combined it leads to a quasi
regular 4D polytope invariant under the Coxeter-Weyl group $W(F_4)$.
 Each vertex of the new polytope is shared by one cube and
three truncated octahedra. Dual of the snub 24 cell is also
constructed. Relevance of these structures to the Coxeter groups  $W(H_4)$ and $W(E_8)$ has been pointed out.
\end{abstract}

\section{INTRODUCTION}
 The $O(4)$ symmetry or rather its proper subgroup  $ SO(4)\approx SU(2)\times SU(2)$ has many applications in physics. A few examples are in order.
 One example is from the atomic physics. The accidental degeneracy of the
energy levels of the hydrogen atom can be explained in terms of the
symmetry of the Hamiltonian under the  $SO(4)$ invariance. Another one is from the condensed matter physics. The
superfluid phases of the  $^3He$ are described by the broken symmetry  $SO(4)\times U(1)$\cite{1}. Other applications can be given from the nuclear physics where the 
symmetry $SU(2)_S\times SU(2)_I$ explains the spin-isospin invariance of the nuclear interactions. In particle physics the same symmetry has been used as a chiral 
symmetry to explain the low masses of the up-down quarks, moreover, it is also suggested as the extension of the electroweak theory. The above applications are all based
 on the Lie group structure of the $O(4)$ symmetry.

 We do not yet have any evidence as to how any finite subgroup
of the group   $O(4)$ has been observed in any physical phenomena. On the other hand the finite subgroups of the group  $O(3)$
 derived from the Coxeter groups  $W(A_3), W(B_3)$ and $W(H_3)$ corresponding respectively to the tetrahedral, octahedral and icosahedral symmetries, have many 
applications in physics, chemistry and biology. The orbits of these groups represent the platonic solids, Archimedean solids as well as the Catalan solids. 
They are all useful in the classifications of the molecules as well as the viral
structures.
 
 The extensions of the polyhedral symmetries to the
4-dimensions can be made by invoking the Coxeter
groups\cite{2}  $W(A_4), W(B_4), W(F_4)$ and $W(H_4)$ which constitute the finite subgroups of the group $O(4)$\cite{3}. However, not all semi regular polytopes can be 
obtained from the above groups. The snub 24-cell is one of the example that requires the proper subgroup of the Coxeter group $W(D_4)$.
 This follows from an observation that the snub icosahedron (chiral
icosahedron) can be derived as the orbit of the proper subgroup of the
tetrahedral group\cite{4}, namely, the orbit  $(W(A_3)/C_2)(\tau,1,\tau)$.
  It is well known that the chiral polyhedra, in particular, snub cube
and snub dodecahedron are used to describe certain molecular
symmetries\cite{5}. This paper studies the construction of
the snub 24-cell by expressing its symmetry group $(W(D_4)/C_2):S_3$
 as well as the vertices of the snub 24-cell in terms of quaternions.
Orbits of the Coxeter-Weyl group $W(D_4)$ are important because they are used to describe the irreducible representations of the $SO(8)$ Lie group which is the little group of the superstring theories in 10
dimensions\cite{6}. Section~\ref{sec:2} deals with the construction
of the group elements from the Coxeter-Dynkin diagram  $D_4$.
 In Section~\ref{sec:3} we construct the chiral icosahedron using the proper
subgroup of the tetrahedral group $W(A_3)$.
Section~\ref{sec:4} shows the explicit construction of the vertices of
two snub 24-cells, mirror images of each other, and explains their cell
structures. The dual polytope of the snub 24-cell has also been
constructed. We point out that the union of two mirror images of these
polytopes leads to a quasi regular polytope with 192 vertices consisting
of the cells of cubes and truncated octahedra. Section~\ref{sec:5} briefs our
method and give some suggestions for the use of snub 24-cell.

\section{CONSTRUCTION OF THE GROUP ${\bf (W(D_4)/C_2):S_3}$ IN TERMS OF QUATERNIONS}\label{sec:2}

 Let  $q=q_0+q_ie_i, (i=1,2,3)$ be a real quaternion with its conjugate defined by $\overline{q}=q_0-q_ie_i$ where the quaternionic imaginary units satisfy the relation

\begin{equation}
e_ie_j=-\delta_{ij}+\epsilon_{ijk}e_k,  (i,j,k=1,2,3). \label{eq:1}
\end{equation}

 Here  $\delta_{ij}$ and $\epsilon_{ijk}$ are the Kronecker and Levi-Civita symbols and summation over the repeated indices is implicit. 
Quaternions generate the four dimensional Euclidean space where the quaternionic scalar product is defined as 

\begin{equation} 
(p,q)=\frac{1}{2}(\overline{p}q+\overline{q}p).\label{eq:2}
\end{equation} 

\begin{figure}[ht]
\begin{center}
\psfig{file=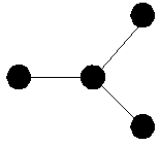,width=0.8752in}
\caption{The Coxeter-Dynkin diagram of $D_4$}\label{fig:1}
\end{center}
\end{figure}

The group of unit quaternions is isomorphic to $SU(2)$ which is the double cover of the proper rotation group  $SO(3)$. 
The quaternionic units can be represented by the Pauli matrices as $e_j=-i\sigma_j, (j=1,2,3)$. In an earlier paper\cite{7} we have constructed 
the group $W(D_4)$ using the quaternionic representations of the simple roots of the diagram  $D_4$.
Here we employ slightly different representation of the group $W(D_4)$ in terms of quaternions. The Coxeter-Dynkin diagram of $D_4$ is shown in Fig.~\ref{fig:1}.

The simple roots are chosen in terms of quaternions as follows:

\begin{equation}
\alpha_1=e_2-e_3, \alpha_2=e_1+e_3, \alpha_3=-e_2-e_3, \alpha_4=1-e_1.\label{eq:3}
\end{equation}

The Cartan matrix  $(C)_{ij}=(\alpha_i, \alpha_j)$ and its inverse  $(C)^{-1}_{ij}=(\omega_i, \omega_j)$, with $(\omega_i, \alpha_j)=\delta_{ij}$, can be written as

\begin{equation}
C=\left(\begin{array}{rrrr}
2 & -1 & 0 & 0 \\
-1 & 2 & -1 & -1 \\
0 & -1 & 2 & 0 \\
0 & -1 & 0 & 2 
\end{array}\right) ,\ \ \  C^{-1}=\frac{1}{2}\left(\begin{array}{cccc}
2 & 2 & 1 & 1 \\
2 & 4 & 2 & 2\\
1 & 2 & 2 & 1 \\
1 & 2 & 1 & 2 \end{array}\right) \label{eq:4}
\end{equation}
 
The weight vectors can be expressed in terms of quaternions as follows:

\begin{equation}
\omega_1=\frac{1}{2}(1+e_1+e_2-e_3), \omega_2=1+e_1, \omega_3=\frac{1}{2}(1+e_1-e_2-e_3), \omega_4=1. \label{eq:5}
\end{equation}
 
For an arbitrary quaternionic root   $\alpha_i$, the reflection generator   $r_i$  can be written as the quaternion 
multiplication  $r_i\Lambda=-\frac{1}{2}\alpha_i\overline{\Lambda}\alpha_i$ simply denoted by 
$r_i=\left[\frac{\alpha_i}{\sqrt{2}},-\frac{\alpha_i}{\sqrt{2}}\right]^*$ as an abstract group element. 
The Coxeter group   $W(D_4)$ generated by the reflection generators  $r_i, (i=1,2,3,4)$ is a group of order 192 
and can be compactly written as 
\begin{equation}
W(D_4)=\{[V_0,V_0]\oplus[V_+,V_-]\oplus[V_-,V_+]\oplus[V_1,V_1]^*\oplus[V_2,V_2]^*\oplus[V_3,V_3]^*\} \label{eq:6}
\end{equation}

where the sets of quaternions are defined by 
\begin{equation}\begin{array}{ll}
V_0=\{\pm 1,\pm e_1,\pm e_2,\pm e_3\}, & V_+=\frac{1}{2}(\pm 1\pm e_1\pm e_2\pm e_3), \\
 & \textnormal{(even number of (-) sign)}\\
 V_-=\frac{1}{2}(\pm 1\pm e_1\pm e_2\pm e_3), & V_1=\{\frac{1}{\sqrt{2}}(\pm 1\pm e_1),\frac{1}{\sqrt{2}}(\pm e_2\pm e_3)\}, \\ 
\textnormal{(odd number of (-) sign)} & \\ \label{eq:7}
 V_2=\{\frac{1}{\sqrt{2}}(\pm 1\pm e_2),\frac{1}{\sqrt{2}}(\pm e_3\pm e_1)\}, & V_3=\{\frac{1}{\sqrt{2}}(\pm 1\pm e_3),\frac{1}{\sqrt{2}}(\pm e_1\pm e_2)\}. 
\end{array}
\end{equation}

 The proper rotation subgroup   $W(D_4)/C_2={[V_0,V_0]\oplus[V_+,V_-]\oplus[V_-,V_+]}$ is simply generated by the set of rotation generators   $r_ir_j, (i,j=1,2,3,4)$.
  This is a group of order 96. With the permutation symmetry  $S_3$ of the simple roots (Dynkin diagram symmetry of the diagram $D_4$) 
the proper rotation subgroup can be extended to the larger group  $(W(D_4)/C_2):S_3$ of order 576 where the notation (:) indicates the semi-direct product.
The generators of the symmetric group  $S_3$ can be taken as  $[\frac{1}{2}(1-e_1+e_2-e_3),\frac{1}{2}(1+e_1+e_2+e_3)]\in[V_+,V_+]$ and $[e_2, -e_2]^*$.
 Then the group can be represented as 
\begin{equation}
(W(D_4)/C_2):S_3=\{[p,q]\oplus[p,q]^*\}, \text{where  } p,q \in T. \label{eq:8}
\end{equation}
 
Here  $T=V_0\oplus V_+\oplus V_-$ represents the quaternionic elements of the binary tetrahedral group of order 24 which also stands for the vertices of the
 polytope 24-cell. We will use a compact notation
\begin{equation}
(W(D_4)/C_2):S_3=\{[T, T]\oplus[T, T]^*\} \label{eq:9}
\end{equation}
for the designation of the group in (\ref{eq:8}). If we had constructed the extension of the group  $W(D_4)$
by the symmetric group   $S_3$ we would  obtain the group\cite{7, 8}, 

\begin{equation}
W(D_4):S_3\approx W(F_4)=\{[T, T]\oplus[T, T]^*\oplus[T', T']\oplus[T', T']^*\} \label{eq:10}
\end{equation}
 where $T'=V_1\oplus V_2\oplus V_3$. We recall that the set $T'$
 represents the vertices of the 24-cell rotated with respect to the set $T$.
It is clear that the group in (\ref{eq:9}) is one of the maximal
subgroups of the group $W(F_4)$ and also it is a maximal subgroup of the Coxeter
group~\cite{9} $W(H_4)=\{[I, I]\oplus[I, I]^*\}$. Here   $I=T\oplus S$ represents the set of 120 quaternions 
representing the binary icosahedral group. As we will show later that the set $S$ with 96 vertices represents
the snub 24 cell.

\section{CHIRAL ICOSAHEDRON CONSTRUCTED AS THE ORBIT OF THE ${\bf W(A_3)/C_2}$}\label{sec:3}

\begin{figure}[ht]
\begin{center}
\psfig{file=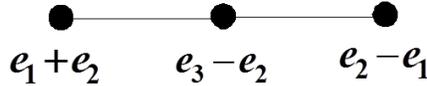,width=2.3in}
\caption{The Coxeter diagram $A_3$ with quaternionic simple roots.}\label{fig:2}
\end{center}
\end{figure}

 Let the Coxeter-Dynkin diagram $A_3$ be represented by the quaternionic simple roots as shown in Fig.~\ref{fig:2}.

The Cartan matrix of the Coxeter diagram $A_3$ and its inverse matrix are given respectively by the matrices

\begin{equation}
C=\left(\begin{array}{rrr}
2 & -1 & 0 \\
-1 & 2 & -1  \\
0 & -1 & 2  
\end{array}\right) ,\ \ \ \ C^{-1}=\frac{1}{4}\left(\begin{array}{ccc}
3 & 2 & 1  \\
2 & 4 & 2 \\
1 & 2 & 3 \end{array}\right). \label{eq:11}
\end{equation}

The basis vectors of the Coxeter-Dynkin diagram $A_3$  in the dual space are given by 

\begin{equation}
\omega'_1=\frac{1}{2}(e_1+e_2+e_3), \  \omega'_2=e_3,\ \omega'_3=\frac{1}{2}(-e_1+e_2+e_3). \label{eq:12}
\end{equation}

Using the simple roots of $A_3$ one can generate the Coxeter group as

\begin{equation}
W(A_3)=\{[T, \overline{T}]\oplus [T', \overline{T'}]^*\}\approx S_4\approx T_d \label{eq:13}
\end{equation}
isomorphic to the tetrahedral group of order 24. The proper rotational
tetrahedral group generated by the rotation generators $r_ir_j, (i,j=1,2,3)$ is given by the set of elements  
$W(A_3)/C_2=[T,\overline{T}]$ of order 12. It has been proven that~\cite{4} one can
generate the vertices of an icosahedron as the orbit of the group  
$W(A_3)/C_2=[T,\overline{T}]$ acting on the either vector $\Lambda_I=\tau\omega'_1 +\omega'_2 +\tau\omega'_3 \equiv(\tau,1,\tau)$
  \ \ \ \ \ or \ \ \ \ \ \   $\Lambda_{II}=\sigma\omega'_1 +\omega'_2 +\sigma\omega'_3 \equiv(\sigma,1,\sigma)$
  \ \ where \ \   $\tau=\frac{1+\sqrt{5}}{2}$
  \ \ \ and \ \ \ \ \   $\sigma=\frac{1-\sqrt{5}}{2}$
  \ as defined in the abstract. \ One can prove that these vectors are
the mirror images of each other. By substituting the quaternionic
expressions in (\ref{eq:11}) \ \ the vectors read \ \ \  $-\sigma\Lambda_I=(e_2+\tau e_3)$ and $\tau^2\Lambda_{II}=(-\tau e_2+e_3)$. 
  Applying the group elements of the group  $W(A_3)/C_2=[T,\overline{T}]$ on these vectors one can generate two sets of icosahedra

\begin{equation}
\begin{array}{c}
(W(A_3)/C_2)(-\sigma\Lambda_I)=\{(\pm e_1\pm\tau e_2), (\pm e_2\pm\tau e_3), (\pm e_3\pm\tau e_1)\}, \\ \\
(W(A_3)/C_2)(-\tau^2\Lambda_{II})=\{(\pm\tau e_1\pm e_2), (\pm\tau e_2\pm e_3), (\pm\tau e_3\pm e_1)\}. \label{eq:14}
\end{array}
\end{equation}

\begin{figure}[ht]
\begin{center}
\psfig{file=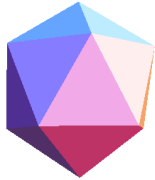,width=0.9827in,height=1.1252in}
\end{center}
\caption{The icosahedron obtained from the orbit $[T,\overline{T}](\tau,1,\tau)$}\label{fig:3}
\end{figure}

The icosahedron represented by the first set of vertices is shown in Fig.~\ref{fig:3}.

 Let us observe that each set is also invariant under the
change of signs of the vectors. This additional symmetry can be
achieved by the operators  $[1,-1]$ and   $[1,1]^*$.
  These generators extend the group to the group   $\{[T,\pm\overline{T}]\oplus[T,\pm\overline{T}]^*\}$ of order 48. Therefore each set is invariant under the above group of order 48. 

One notes that the Dynkin diagram symmetry of the diagram $A_3$, namely,   $\alpha_1\leftrightarrow\alpha_2, \alpha_3\rightarrow\alpha_3$, which leads to the transformation of quaternionic units  $e_1\rightarrow-e_1, e_2\rightarrow e_2$ and $e_3\rightarrow e_3$. This can be achieved by the operator  $[e_1, -e_1]^*$.
  The proper tetrahedral group can be extended by this generator to the
group   $(W(A_3)/C_2):C_2=\{[T,\overline{T}]\oplus[T,\overline{T}]^*\}$.
 Then the above group of order 48 is the direct product of the group
generated by the generator $[1,-1]$ and the latter group. Then it can be written as 
\begin{equation}
[(W(A_3)/C_2):C_2]\times C_2=\{[T,\pm\overline{T}]\oplus[T,\pm\overline{T}]^*\}. \label{eq:15}
\end{equation}

 Here we also note that the group in (\ref{eq:15}) involves the pyritohedral
group  $T_h\approx (W(A_3)/C_2)\times C_2=\{[T,\pm\overline{T}]\}$  as a maximal subgroup. The pyritohedral group represents the symmetry
of the iron pyrites.

\begin{figure}
\begin{center}
\psfig{file=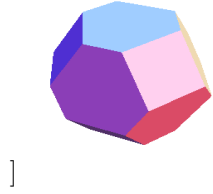,width=1.25in,height=1.1854in]}
\caption{The quasi regular polyhedron represented by the vertices of (\ref{eq:14})}\label{fig:4}
\end{center}
\end{figure}

 The second set in (\ref{eq:14}) represents the mirror image of the icosahedron
depicted in Fig.~\ref{fig:3}. The generation of the vertices of an icosahedron from the proper tetrahedral group is interesting which will help us to construct the snub 24-cell from the diagram  $D_4$ as the orbit of $(W(D_4)/C_2):S_3=\{[T,T]\oplus[T,T]^*\}$. We also note an interesting observation that when two sets of orbits
in (\ref{eq:14}) are combined, it represents a solid with 24 vertices. The
emerging polyhedron is a quasi regular truncated octahedron obtained as
the orbit~\cite{4}  $W(B_3)(1,\tau,0)$.
  Here the group  $W(B_3)$ is the octahedral group generated by reflections from the Coxeter
diagram   $B_3$.
  The quasi regular truncated octahedron is shown in Fig.~\ref{fig:4}. Here the
faces of the solid are squares of side  $\tau$
  and the isogonal hexagons with edge lengths  $\tau$ and $1$.
 
\section{CONSTRUCTION OF SNUB 24-CELL FROM THE DIAGRAM ${\bf D_4}$}\label{sec:4}

 It is clear from the Coxeter-Dynkin diagram of Fig.~\ref{fig:1} that one can generate three tetrahedral subgroups by using the sets of generators $<r_1,r_2,r_3>, <r_3, r_2, r_4>, <r_4, r_2, r_1>$
  and their proper tetrahedral subgroups are generated by the sets  
$<r_1r_2, r_2r_3, r_3r_1>, <r_3r_2, r_2r_4, r_4r_3>, <r_4r_2, r_2r_1, r_1r_4>$. In terms of quaternionic notation they can be written as 

\begin{equation}
\begin{array}{l}
<r_1r_2, r_2r_3, r_3r_1>=[T,\overline{\omega_4}\overline{T}\omega_4], \\
<r_3r_2, r_2r_4, r_4r_3>=[T, \overline{\omega_1}\overline{T}\omega_1], \\
<r_4r_2, r_2r_1, r_1r_4>=[T, \overline{\omega_3}\overline{T}\omega_3]. 
\end{array} \label{eq:16}
\end{equation}
 
 Recall the definition of the weight vectors in terms of
quaternionic units in (\ref{eq:5}). These proper tetrahedral groups leave the
vectors   $\omega_4, \omega_1$ and $\omega_3$ invariant respectively. Actually the pyritohedral group   $(W(A_3)/C_2)\times C_2=[T,\pm\overline{T}]$ can be embedded 
in the group $W(D_4)/C_2$ in 12 different ways, in each case, the quaternion  $\pm q\in T$ will be left invariant. One can generalize the idea discussed in
Section~\ref{sec:3} to the case of the group  $W(D_4)/C_2$. It is then straightforward to show that the vectors $\Lambda_I=\tau(\omega_1+\omega_3+\omega_4)+\omega_2$
 and  $\Lambda_{II}=\sigma(\omega_1+\omega_3+\omega_4)+\omega_2$ can be used to generate two sets of snub 24 cells which are mirror images of each other provided 
suitable scale factors are chosen. In terms of quaternions these vectors read

\begin{equation}
\Lambda'_I=\frac{1}{2}\sigma^2\Lambda_I=\frac{1}{2}(\tau+e_1+\sigma e_3), \Lambda'_{II}=\frac{1}{2}\tau^2\Lambda_{II}=\frac{1}{2}(\sigma+e_1+\tau e_3). \label{eq:17}
\end{equation} 
 Note that these vectors are transformed to each other under the exchange of  $\tau\leftrightarrow\sigma$. 
 One can generate the vertices of the snub 24-cell by applying the group
 $W(D_4)/C_2$ on these vectors.

 By applying the group elements  $(W(D_4)/C_2):S_3=\{[T,T]\oplus[T,T]^*\}$ on the vectors in (\ref{eq:17}) one can generate two sets of 96 elements each. 
For example, $ [(W(D_3)/C_2):S_3]$ leads to the following set consisting of 96 unit quaternions~\cite{9}

\begin{equation}
\begin{array}{rcl}
S=\{\frac{1}{2}(\pm \tau\pm e_1\pm\sigma e_3),& \frac{1}{2}(\pm \tau\pm e_2\pm\sigma e_1), & \frac{1}{2}(\pm \tau\pm e_3\pm\sigma e_2), \\ \\
\frac{1}{2}(\pm \sigma\pm e_1\pm\tau e_2),& \frac{1}{2}(\pm \sigma\pm e_2\pm\tau e_3),&\frac{1}{2}(\pm \sigma\pm e_3\pm\tau e_1), \\ \\
\frac{1}{2}(\pm 1 \pm\tau e_1\pm\sigma e_2),& \frac{1}{2}(\pm 1 \pm\tau e_2\pm\sigma e_3),& \frac{1}{2}(\pm 1 \pm\tau e_3\pm\sigma e_1), \\ \\
\frac{1}{2}(\pm\sigma e_1 \pm\tau e_2\pm e_3),& \frac{1}{2}(\pm\sigma e_2 \pm\tau e_3\pm e_1),& \frac{1}{2}(\pm\sigma e_3 \pm\tau e_1\pm e_2)\}. \label{eq:18}
\end{array}
\end{equation}
 
 Let us recall that the set of 120 unit quaternions $I=T\oplus S$ represents the elements of the binary icosahedral group as well as the
vertices of the Platonic polytope 600-cell~\cite{10}
consisting of 600 tetrahedra, the symmetry of which is the Coxeter
group   $W(H_4)=\{[I, I]\oplus[I,I]^*\}$. Therefore the set $I=T\oplus S$ represents the decomposition of the  $W(H_4)$ polytope under its maximal subgroup  $(W(D_4)/C_2):S_3$  which branches into the set of 24-cell and the set of snub 24-cell.
Applying the same group on the second vector $\frac{1}{2}\tau^2\Lambda_{II}=\frac{1}{2}(\sigma+e_1+\tau e_3)$ we would generate the set  
$\tilde{S}=S(\tau\leftrightarrow\sigma) $ which will be the mirror image of the set  $S$ in (\ref{eq:18}), that is, one can show for example that $r_1S=\tilde{S}$.
 Of course, this is true for any reflection generators of the group  
$W(D_4)$. The operation $(\sim)$ exchanges $\tau\leftrightarrow\sigma$.
  So here it plays the role of reflection symmetry for the two mirror
images of the snub 24-cell. It is well known that the binary
icosahedral group\cite{11} has two 2-dimensional irreducible
representations; if one is represented by the set $I=T\oplus S$
 the other is represented by the set $\tilde{I}=T\oplus \tilde{S}$.
  Two irreducible representations of the binary icosahedral group will
lead to two 4-dimensional irreducible representations of the Coxeter
group in the form   $W(H_4)=\{[I,I]\oplus[I,I]^*\}$ and $W(H_4)=\{[\tilde{I},\tilde{I}]\oplus[\tilde{I},\tilde{I}]^*\}$.
  Their common maximal subgroup is $(W(D_4)/C_2):S_3=\{[T,T]\oplus[T,T]^*\}$. Since $W(H_4)$
 is a maximal subgroup of the Coxeter-Weyl group $W(E_8)$\cite{12} and the root system of the $E_8$
 can be represented as the union of the quaternions  $I\oplus\sigma I$ or $\tilde{I}\oplus\sigma\tilde{I}$\cite{13}, this analogy can be traced back to the
Coxeter-Weyl group   $W(E_8\oplus E_8)$ and perhaps to the heterotic superstring theory.

 Now we come back to the structure of the snub 24-cell. In view
of the arguments raised in Section~\ref{sec:3}, the group structure in (\ref{eq:15})
can be extended to the groups

\begin{equation}
[T,\overline{\omega_4}\overline{T}\omega_4]\oplus[T,\omega_4\overline{T}\omega_4]^*, [T,\overline{\omega_1}\overline{T}\omega_1]\oplus[T,\omega_1\overline{T}\omega_1]^*, 
[T,\overline{\omega_3}\overline{T}\omega_3]\oplus[T,\omega_3\overline{T}\omega_3]^*. \label{eq:19}
\end{equation}
 
Each group acting on the vector  $\Lambda'_I$ will generate an icosahedron. The three icosahedra share the same
vertex  $\frac{1}{2}(\tau+e_1+\sigma e_3)$.  We can write the vertices of these icosahedra in terms of quaternions
from the set  $S$ but it is not needed at this moment. But we note an important aspect that the centers of these three icosahedra can be represented by the vectors   $(\omega_4, \omega_1, \omega_3)\in T$
 respectively up to some scale factor. We can prove that the following
sets of vectors 

\begin{equation}
\begin{array}{cc}
P(1)=(\Lambda'_I, r_1r_3\Lambda'_I,r_3r_4\Lambda'_I, r_4r_1\Lambda'_I), & P(2)=(\Lambda'_I, r_2r_1\Lambda'_I,r_2r_3\Lambda'_I, r_2r_4\Lambda'_I), \\
P(3)=(\Lambda'_I, r_3r_2\Lambda'_I,r_3r_1\Lambda'_I, r_3r_4\Lambda'_I), & P(4)=(\Lambda'_I, r_4r_2\Lambda'_I,r_4r_3\Lambda'_I, r_4r_1\Lambda'_I), \\
P(5)=(\Lambda'_I, r_1r_2\Lambda'_I,r_1r_1\Lambda'_I, r_1r_4\Lambda'_I), & \label{eq:20}
\end{array}
\end{equation}
form five tetrahedra. Therefore the vector  $\Lambda'_I$ is shared by three icosahedra and five tetrahedra. We note that the
Dynkin diagram symmetry  $S_3$ permutes the generators $(r_1, r_3, r_4)$ but leaves  $r_2$ invariant. Similarly it permutes the vectors  
 $(\omega_1, \omega_3, \omega_4)$ and leaves the vector $\omega_2$ invariant and consequently it leaves the vector $\Lambda'_I$ invariant. 
This shows that the two tetrahedra  $P(1)$ and $P(2)$ are left invariant by the permutation symmetry while the tetrahedra  $P(3), P(4)$ and $P(5)$ are 
permuted by the symmetric group  $S_3$. One notes that the group generated by  $<r_1r_3, r_1r_4>\approx C_2\times C_2$ (Klein four-group) can be extended by 
the permutation group   $S_3$ to the tetrahedral group  $T_d\approx (C_2\times C_2):S_3=\{[T,\overline{\omega_2}\overline{T}\omega_2]\oplus[T,\omega_2\overline{T}\omega_2]^*\}$ 
leaving the vector  $\omega_2$ invariant which represents the center, up to a scale factor, of the tetrahedron  $P(1)$.
 
 This group can be extended to the group  
$T_d\times C_2\approx[(C_2\times C_2):S_3]\times C_2=\{[T,\pm\overline{\omega_2}\overline{T}\omega_2]\oplus[T,\pm\omega_2\overline{T}\omega_2]^*\}$ 
which leaves the vectors  $\pm w_2$ invariant. Then the group   $T_d\times C_2$ leaving the vectors  
$\pm\frac{\omega_2}{\sqrt{2}}=\pm\frac{1}{\sqrt{2}}(1+e_1)\in T'$ invariant can be embedded in the group $(W(D_4)/C_2):S_3=\{[T,T]\oplus[T, T]^*\}$ 
in 12 different ways. This indicates that the snub 24 cell involves 24
tetrahedra each left invariant under one of the conjugate tetrahedral group  $T_d$.
 The centers of these 24 tetrahedra constitute the vertices of a 24-cell $T'$ up to some scale factor. Similarly the symmetry group of icosahedrons 
 $[T,\pm \overline{q}\overline{T}q]\oplus[T,\pm q\overline{T}q]^*, \pm q\in T$ can be embedded in the group   $\{[T, T]\oplus[T, T]^*\}$ in 12 different ways where  $\pm q\in T$ 
left invariant. Actually it follows that the centers of the 24 icosahedra belong to the sets  $V_0, V_+, V_-$ so that the icosahedra are grouped as 24=8+8+8. 
Indeed the sets of
vectors   $V_0, V_+, V_-$ constitute the weights of the 8-dimensional irreducible representations $8_v, 8_s, 8_c$ of the Lie group $SO(8)$.
  Now what remains is the other set of tetrahedra. One can show that the
remaining sets of tetrahedra possess the permutation symmetry  
$S_3$ only. The number of tetrahedra possessing the $S_3$ symmetry is  

$$
\frac{\vert(W(D_4)/C_2):S_3\vert}{\vert S_3\vert}=\frac{576}{6}=96. \label{eq:21}
$$ Therefore the total number of cells of the snub 24-cell is 144 which
consists of 24 icosahedra, 24 tetrahedra and 96 tetrahedra. 

 The dual of the snub 24-cell has been constructed in the
reference\cite{9}. To construct the dual polytope of the
snub 24 cell one should determine the vectors representing the centers
of the cells (here three icosahedra and five tetrahedra) sharing the
same vertex. We have already listed the vectors representing the
centers of the icosahedra. The vectors, up to some scale factors
representing the centers of the tetrahedra can be written as 

\begin{equation}
\begin{array}{l}
P'(1)=\omega_2,\\ \\
P'(2)=\tau^4[\tau(\omega_1+\omega_3+\omega_4)-\omega_2], \\ \\
P'(3)=\tau^3[(\omega_1-\omega_3+\omega_4)+\tau^2\omega_2], \\ \\
P'(4)=\tau^3[(\omega_1+\omega_3-\omega_4)+\tau^2\omega_2], \\ \\
P'(5)=\tau^3[(-\omega_1+\omega_3+\omega_4)+\tau^2\omega_2]. \label{eq:22}
\end{array}
\end{equation}

 It is clear that while the first two vectors remain invariant
under the symmetry   $S_3\approx D_3$  the last three vectors are permuted to each other. The last three
vectors   $P'(3), P'(4), P'(5)$ form an equilateral triangle parallel to the equilateral triangle formed by the unit vectors  $(\omega_1, \omega_3, \omega_4)$. 
It is better to express the other five vectors also as unit quaternions using the expressions in (\ref{eq:5})

\begin{equation}
\begin{array}{l}
c(1)=\frac{1}{\sqrt{2}}P'(1)=\frac{1}{\sqrt{2}}(1+e_1), \\ \\
c(2)=\frac{\sigma^4}{2\sqrt{2}}P'(2)=\frac{1}{2\sqrt{2}}[(\tau-\sigma)-\tau e_1-\tau e_3], \\ \\
c(3)=\frac{\sigma^4}{2\sqrt{2}}P'(3)=\frac{1}{2\sqrt{2}}[(\tau-\sigma)+\tau e_1-\sigma e_2], \\ \\
c(4)=\frac{\sigma^4}{2\sqrt{2}}P'(4)=\frac{1}{2\sqrt{2}}[\tau+(\tau-\sigma)e_1+\sigma e_3], \\ \\
c(5)=\frac{\sigma^4}{2\sqrt{2}}P'(5)=\frac{1}{2\sqrt{2}}[(\tau-\sigma)+\tau e_1+\sigma e_2]. \label{eq:23}
\end{array}
\end{equation}

 Note that the vector $c(1)-c(2)$  is orthogonal to the planes determined by the sets of vectors  
$(\omega_1, \omega_3, \omega_4)$ and  $(c_3, c_4, c_5)$. To determine the structure of the cell of the dual snub 24-cell one
needs to express these eight vectors in the hyperplane orthogonal to
the vertex   $\Lambda'_I=\frac{1}{2}(\tau+e_1 +\sigma e_3)$. We have already noted that the plane determined by   $(\omega_1, \omega_3, \omega_4)$ 
is orthogonal to the vector  $\Lambda'_I$. One can also check that the hyperplane determined by the five vectors in
(\ref{eq:23}) are also orthogonal to the vector $\Lambda'_I$.  However, the planes determined by the set of vectors in (\ref{eq:23}) and the
vectors $(\omega_1, \omega_3, \omega_4)$ are not orthogonal to the vector  $\Lambda'_I$ unless a change of relative scale factor between these sets of vectors is introduced. 
Indeed the set of eight vectors  $[\frac{\tau}{\sqrt{2}}\omega_1, \frac{\tau}{\sqrt{2}}\omega_3, \frac{\tau}{\sqrt{2}}\omega_4, c_1, c_2, c_3, c_4, c_5]$ forms an 
hyperplane orthogonal to the vertex $\Lambda'_I$.
  Now one can express above eight vectors in a new basis of quaternions
defined by  $(p_0\equiv\Lambda'_I, p_1=e_1p_0, p_2=e_2p_0, p_3=e_3p_0)$
 then all eight vectors will have the same component along the unit
vector  $p_0$. Then the rest of the components define the vectors in the space orthogonal to the vector  $p_0$ which reads
\begin{equation}
\begin{array}{l}
\frac{1}{2\sqrt{2}}(-\tau,0,1),\ \frac{1}{2\sqrt{2}}(0,-1,-\tau),\ \frac{1}{2\sqrt{2}}(1,\tau,0),\ \frac{1}{2\sqrt{2}}(-\sigma,\sigma,-\sigma),\ \frac{1}{2\sqrt{2}}(\sigma,-\sigma,\sigma),\\ \\
\frac{1}{2\sqrt{2}}(\sigma^2,0,1),\ \frac{1}{2\sqrt{2}}(1,-\sigma^2,0),\ \frac{1}{2\sqrt{2}}(0,1,\sigma^2). \label{eq:24}
\end{array}
\end{equation}

\begin{figure}[ht]
\begin{center}$
\begin{array}{cc}
\psfig{file=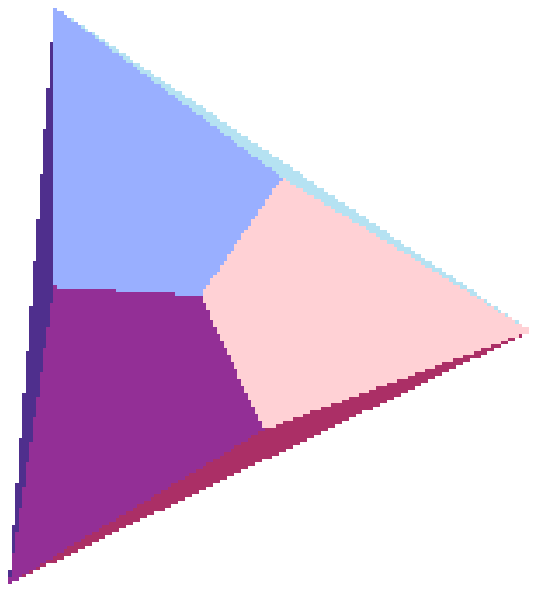,width=1.1126in,height=1.1874in}
&
\psfig{file=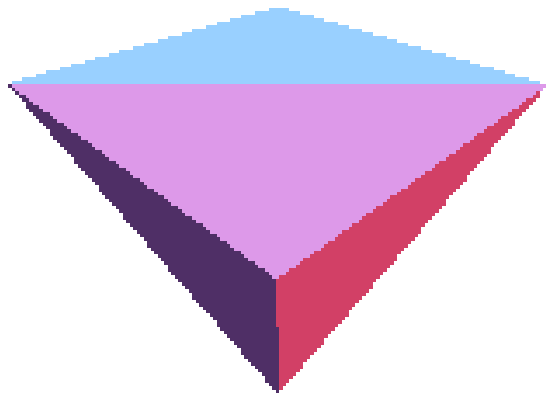,width=1.4146in,height=1.098in}
\end{array}$
\end{center}
\caption{Typical cell of the dual polytope of the snub 24 cell. (a) Top view (b) Bottom view}\label{fig:5}
\end{figure}

 This solid consists of 9 faces. Three faces are made of kites of sides 
 $\frac{\sigma^2}{\sqrt{2}}$ and $\frac{1}{\sqrt{2}}$
  and six triangular faces of the form of isosceles triangles of sides  
$\frac{\tau}{\sqrt{2}}$ and $\frac{1}{\sqrt{2}}$.
  It has a dihedral symmetry  $D_3$.
  This typical cell of the dual polytope is depicted in Fig.~\ref{fig:5}.

\begin{figure}[ht]
\begin{center}
\psfig{file=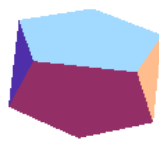,width=1.3854in,height=1.3016in}
\caption{Tridiminished icosahedron}\label{fig:6}
\end{center}
\end{figure}

 The vertex figure of any convex polytope is the convex solid
formed by the nearest vertices to the vertex. The following nine
vertices constitute the set of nearest vectors to the vector  $\Lambda'_I$
\begin{equation}
r_1r_2\Lambda'_I, r_2r_1\Lambda'_I, r_1r_3\Lambda'_I, r_2r_3\Lambda'_I, r_3r_2\Lambda'_I, r_1r_4\Lambda'_I, r_4r_2\Lambda'_I, r_2r_4\Lambda'_I, r_3r_4\Lambda'_I. \label{eq:25}
\end{equation} 
When they are expressed in the quaternionic bases $ (p_0\equiv\Lambda'_I, p_1=e_1p_0, p_2=e_2p_0, p_3=e_3p_0)$
  and the components of   $p_0$ are removed the nine vertices read
\begin{equation}
(\pm 1,0,\sigma),(1,0,-\sigma),(\sigma,\pm 1,0), (0,\sigma,1), (-\sigma,-1,0), (0,-\sigma,\pm 1). \label{eq:26}
\end{equation}
 These vertices form the solid called tridiminished icosahedrons, one
of the Johnson's solid, $J_{63}$\cite{14} as shown in Fig.~\ref{fig:6}. If the additional three
vectors   $(-1,0,-\sigma),(\sigma, 1,0), (0,\sigma,-1)$
 are combined with the vectors in (\ref{eq:26})  we would obtain a regular
icosahedron.
 Similar to what we have done for the two mirror images of the
icosahedra in Section~\ref{sec:3},  we can combine two mirror images of the
snub 24 cell namely we take the union of the sets   $S$ and $\tilde{S}$
to form the set of vectors   $S\oplus\tilde{S}$ which is now reflection symmetric. We observe that the symmetry of the
new system extends to the group represented by the Coxeter-Weyl group  
$W(D_4):S_3\approx W(F_4)$ given in (\ref{eq:10}). It can be shown that the 192 vectors of the set  $S\oplus\tilde{S}$ can be obtained from the vector  $\Lambda=\tau\omega_3+\omega_4\equiv(0,0,\tau,1)$ as the orbit   $W(F_4)\Lambda$. To obtain exactly the same set of elements of  $S\oplus\tilde{S}$, we choose the quaternionic simple roots of $W(F_4)$ as 
\begin{equation}
\alpha_1=\frac{1}{\sqrt{2}}(1-e_1-e_2-e_3), \alpha_2=\sqrt{2}e_3, \alpha_3=e_2-e_3, \alpha_4=e_1-e_2 \label{eq:27}
\end{equation}
which define the Cartan matrix and its inverse as follows:

\begin{equation}
C_{F_4}=\left(\begin{array}{rrrr} 2 & -1 & 0 & 0 \\ -1 &2 &-\sqrt{2} & 0 \\ 0 & -\sqrt{2} & 2 & -1 \\ 0 & 0 & -1 & 2 
\end{array}
\right), \ \ (C_{F_4})^{-1}=\left(\begin{array}{cccc}
2 & 3 & 2\sqrt{2} & \sqrt{2} \\ 3 & 6 & 4\sqrt{2} & 2\sqrt{2} \\ 2\sqrt{2}& 4\sqrt{2} &6 & 3 \\ \sqrt{2} & 2\sqrt{2} & 3 & 2
\end{array}\right). \label{eq:28}
\end{equation}

One can readily show that 

\begin{equation}
\Lambda'\equiv\frac{\sigma^2}{2}\Lambda\equiv\frac{\sigma^2}{2}(0,0,\tau,1)=\frac{1}{2}(\tau+e_1-\sigma e_2)\in \tilde{S}. \label{eq:29}
\end{equation}
 
Applying the group elements of the Coxeter group $W(F_4)$ one generates the set of vectors of   $S\oplus\tilde{S}$. The orbit  $\frac{\sigma^2}{2}W(F_4)(0,0,\tau,1)$ 
has an interesting structure. It is a polytope consisting of two types of cells; cube and truncated octahedron shown in the Fig.~\ref{fig:4}. 
To analyze the cell structures of the orbit $W(F_4)(0,0,\tau,1)$ one needs to look at the subgroups of rank-3. We have two $W(B_3)$ groups generated by respectively 
$W(B_3):\langle r_1,r_2,r_3\rangle$ and $W(B_3):\langle r_2,r_3,r_4\rangle$. The first group acting on the vector  $(0,0,\tau,1)$ generate a cube with sides of  
 $\sqrt{2}\tau$ and the second group will generate a truncated octahedron with faces of isogonal hexagons of edge lengths  $\sqrt{2}\tau$ and $\sqrt{2}$ and a 
square of sides  $\sqrt{2}\tau$.
  The total number of cells of the polytope is 48
(24(cube)+24(quasiregular truncated octahedron)). At each vertex of the
polytope   $W(F_4)(0,0,\tau,1)$  there is one cube and three quasi regular truncated octahedra. To show the validity of our argument we apply the group   
$W(B_3):\langle r_1,r_2,r_3\rangle$  on the vector   $\Lambda'=\frac{1}{2}(\tau+e_1-\sigma e_2)$. This octahedral subgroup is expressed in terms of quaternions as 

\begin{equation}
W(B_3)=[T,\overline{\omega}_4\overline{T}\omega_4]\oplus[T,\omega_4\overline{T}\omega_4]^*\oplus[T',\overline{\omega}_4\overline{T}'\omega]\oplus[T',\omega_4\overline{T}'\omega_4]^* \label{eq:30}
\end{equation}
where  $\omega_4=1+e_1$ which is left invariant. The orbit $W(B_3)\Lambda'=\langle r_1,r_2,r_3\rangle\Lambda'$ will lead to 8 vertices of the cube

\begin{equation}
\begin{array}{c}
\frac{1}{2}(\tau+e_1-\sigma e_2), \frac{1}{2}(\tau+e_1+\sigma e_2),\frac{1}{2}(1+\tau e_1-\sigma e_3), \frac{1}{2}(1+\tau e_1+\sigma e_3), \\ \\
\frac{1}{2}(\tau+e_1-\sigma e_3), \frac{1}{2}(\tau+e_1+\sigma e_3),\frac{1}{2}(1+\tau e_1-\sigma e_2), \frac{1}{2}(1+\tau e_1+\sigma e_2). \label{eq:31}
\end{array}
\end{equation}

To show that the vectors in (\ref{eq:31}) represent the vertices of a cube we
note that the center of the cube is represented by the unit quaternion 
 $p_0=\frac{1}{\sqrt{2}}(1+e_1)$ and one defines a new basis of unit vectors as

\begin{equation}
\begin{array}{lr}
p_0=\frac{1}{\sqrt{2}}(1+e_1), & p_1=e_1 p_0=\frac{1}{\sqrt{2}}(-1+e_1), \\
p_2=e_2 p_0=\frac{1}{\sqrt{2}}(e_2-e_3), & p_3=e_3 p_0=\frac{1}{\sqrt{2}}(e_2+e_3). \label{eq:32}
\end{array}
\end{equation}
 
Then removing the common components of the unit vector  $\frac{\tau^2}{2\sqrt{2}}p_0 $ and ignoring an overall factor  $\frac{\sigma}{2\sqrt{2}}$ the vectors in (\ref{eq:31}) 
would read in the new basis of unit vectors $p_1, p_2, p_3$
 
\begin{equation}
\begin{array}{c}
(1,-1,1),\ (1,1,-1_),\ (-1,-1,-1),\ (-1,1,1), \\
(1,-1,-1),\ (-1,1,-1),\ (1,1,1),\ (-1,-1,1). \label{eq:33}
\end{array}
\end{equation}
 
These represent the vertices of a cube with the first and the second
sets representing two dual tetrahedra embedded in the cube.
 
\begin{figure}[ht]
\begin{center}
\psfig{file=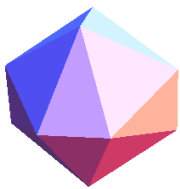,width=1.1783in,height=1.2598in}
\caption{Quasi regular icosahedron with pyritohedral symmetry}\label{fig:7}
\end{center}
\end{figure}

 One can also project the snub 24 cell to 3-dimensional space.
To do this one should choose one of its maximal rank-3 subgroups. As we
noted earlier that the icosahedron in the snub 24 cell has a maximal
pyritohedral symmetry either represented by   $T_h\approx[T,\overline{T}]\oplus[T,-\overline{T}$] or  $T_h\approx[T,\overline{T}]\oplus[T,\overline{T}]^*$ which leads to the same 
result in 3D. Therefore projecting the snub 24 cell into 3D leads to a decomposition of 96 vertices of (\ref{eq:18}) as the orbits of the pyritohedral group $T_h$. 
This leads to 4 icosahedra as shown in Fig.~\ref{fig:3} classified with respect to the real parts $\pm\frac{\tau}{2}$ and $\pm\frac{\sigma}{2}$ of the sets of quaternions in (\ref{eq:18}). 
The other two orbits of size 12 are obtained from the set of quaternions in (\ref{eq:18}) with the real parts given by  $\pm\frac{1}{2}$.
 They lead to the solid in Fig.~\ref{fig:7} which is a quasi regular icosahedron, dual of a pyritohedron~\cite{4}.

\begin{figure}[ht]
\begin{center}
\psfig{file=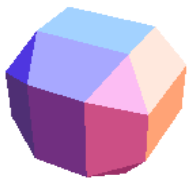,width=1.2264in,height=1.2508in}
\caption{The quasiregular polyhedron with 24 vertices with pyritohedral symmetry}\label{fig:8}
\end{center}
\end{figure}

 The set of 24 quaternions with zero real part constitutes a
single orbit under the pyritohedral group which is plotted in Fig.~\ref{fig:8}.
It consists of three types of faces, equilateral triangle, golden
rectangle and a trapezoid constructed with two edge lengths $2$ and $2\tau^{-1}$.

\section{CONCLUSION }\label{sec:5}

 \ \ The snub 24 cell is an interesting semi regular polytope in
4D. Its cells consist of 24 icosahedra and 120 tetrahedra. Its symmetry
is generated by the proper rotational elements of the Coxeter group  
$W(D_4)$  extended by the Coxeter-Dynkin diagram symmetry.  It exists in two chiral forms represented by the sets  $S$ and $\tilde{S}$ as explained in the text. The union  $S\oplus\tilde{S}$ with 192 vertices forms a quasi regular polytope of the Coxeter-Weyl
group   $W(F_4)$.
  The sets $I=T\oplus S$ and $\tilde{I}=T\oplus\tilde{S}$ are two representations of 600 cell consisting of tetrahedra which are
the orbits of the Coxeter group  $W(H_4)$.
  They can be traced back to the Coxeter group  $W(E_8\oplus E_8)$
  which could be relevant to the heterotic superstring theory.

 The rank-3 subgroups of the symmetry group  $(W(D_4)/C_2):S_3=\{[T,T]\oplus[T,T]^*\}$ of the snub 24-cell are the pyritohedral group $T_h$ and the tetrahedral group  $T_d$ each of order 24. They are the respective symmetry groups of the 24
icosahedra and 24 tetrahedra in the snub 24 cell. Each of the remaining
96 tetrahedral cells possesses permutation symmetry  $S_3$. The dual polytope of the snub 24 cell consists of 144 vertices and 96
cells. Its typical cell is depicted in Fig.~\ref{fig:5}. It is expected that such
a unique chiral polytope with its magnificent symmetry might have some
relevance to the supersymmetric theories in higher dimensions.


\end{document}